\documentclass[twocolumn,showpacs,preprintnumbers,amsmath,amssymb,superscriptaddress,prl]{revtex4}
\pdfoutput=1

\usepackage{dcolumn}
\usepackage{bm}
\usepackage[latin1]{inputenc}
\usepackage{float}
\usepackage[latin1]{inputenc}
\usepackage[pdftex]{graphicx}

\begin{document}

\title{Anisotropy and controllable band structure in supra-wavelength polaritonic metasurfaces}
\author{K. Chevrier}
\author{J.M. Benoit}
\author{C. Symonds}
\affiliation{Univ Lyon, Universit\'e Claude Bernard Lyon 1, CNRS, Institut Lumière Matière, F-69622, LYON, France.}
\author{S.K. Saikin}
\affiliation{Department of Chemistry and Chemical Biology, Harvard University, Cambridge, MA 02138, USA.}
\author{J. Yuen-Zhou}
\affiliation{Department of Chemistry and Biochemistry, University of California San Diego, La Jolla, CA, USA.}
\author{J. Bellessa}
\email{joel.bellessa@univ-lyon1.fr} \affiliation{Univ Lyon, Universit\'e Claude Bernard Lyon 1, CNRS, Institut Lumière Matière, F-69622, LYON, France.}

\date{\today}

\begin{abstract}

In this letter we exploit the extended coherence length of mixed plasmon/exciton states to generate active metasurfaces. For this purpose, periodic stripes of organic dye are deposited on a continuous silver film. Typical metasurface effects, such as effective behavior and geometry sensitivity, are measured for periods exceeding the polaritonic wavelength by more than one order of magnitude. By adjusting the metasurface geometry, anisotropy, modified band structure and unidimensional polaritons are computationally simulated and experimentally observed in reflectometry as well as in emission.

\end{abstract}

\pacs{ 71.36.+c, 42.25.Kb, 73.20.Mf, 78.55.Kz  }
\keywords{}

\maketitle
The strong-coupling phenomenon that occurs when light-matter interaction prevails over damping\cite{Kimble} leads to remarkable changes in the optical properties of matter, modifying both the spectrum\cite{Weisbuch} and dynamics of excitations\cite{Norris}. In molecular systems\cite{Lidzey}, this regime can be reached with a reduced number of molecules, opening the way to the realization of single photon emitters and quantum information processing architectures\cite{Chikkaraddy}. When a large number of emitters is involved, a coherent coupling of the emitters through the electromagnetic mode occurs\cite{Agranovich,Aberra}. In metal/semiconductors systems, the strongly-coupled state (polariton) arising between molecules and a propagating plasmon has an extension ranging from a few\cite{Aberra,Shi} to several tens of microns\cite{Chevrier}. The conductivity of molecular films can be dramatically modified through this coherent coupling\cite{Orgiu,Feist}. The strong-coupling regime also affects other properties such as chemical reactivity\cite{Hutchinson} or efficiency of second harmonic generation\cite{Chervy}. Within this context, the control of polaritonic modes is of crucial interest. Polaritons can be tailored by shaping their electromagnetic part: mesa\cite{Daifa}, pillar structuration\cite{Bajoni} or photonic metasurfaces\cite{Delteil,Benz} have, for example, allowed for an efficient control of their spectra. Various types of plasmonic modes have also been exploited in strong coupling to reduce volume and losses\cite{Fofang,Kleemann}. This control can also be achieved by tailoring the excitonic part: a recent demonstration of polarization control and energy tunability has been performed with aligned carbon nanotubes coupled to a microcavity\cite{Gao}. Beside the properties inherited from their light or matter components, the hybrid nature of polaritonic states leads to unique properties which could be obtained neither by the photon nor exciton alone. This has yield for instance the first observation of Bose condensation in condensed matter physics, with exciton/photon polaritons \cite{Kasprzak,Christopoulos,Kena}. This hybrid nature can also be exploited to tailor the polaritons properties in the linear regime.

In the present work, we exploit the mixing of propagating plasmons with a set of localized emitters to control their band structure and generate anisotropy. More specifically, by patterning an organic film on silver, we investigate a new type of metasurface that builds on the polariton extended coherence. As the coherence length is one order of magnitude larger than the wavelength, an efficient geometry control can be achieved in the visible range with micrometric structuration. Employing this method, we demonstrate anisotropy generation in transition energies and emission patterns. These findings pave the way towards a novel type of artificial materials which require micro- instead of nano-structuration. The developments of new functionalities based on polaritonic metasurfaces could also build on the extraordinary achievements of metamaterials\cite{Engheta,Monticone,Hess} and metasurfaces\cite{Meinzer,Yu}, with potential electrical or optical control enabled by the excitonic part of the polaritons\cite{Anappara,Vasa}.

The sample consists of stripes of active dye, periodically deposited on a 50 nm continuous silver film thermally evaporated on a glass substrate. For this purpose a solution of J-aggregated dye TDBC (5,6-Dichloro-2-[[5,6-dichloro-1-ethyl-3-(4-sulfobutyl)-benzimidazol-2-ylidene]-propenyl]-1-ethyl-3-(4-sulfobutyl)-benzimidazolium hydroxide, inner salt, sodium salt) is spin coated on top of the silver film. The spin coated layer has a thickness of 17 nm, measured by plasmon fitting, with a sharp absorption peak at 2.1 eV. The inactive stripes have been created in the homogeneous cyanine dye film by photobleaching the dye molecules using local UV-irradiation performed with a 355 nm solid state laser. By running an adjustable mechanical slit at an intermediate image plane, a micrometric precision is reachable for the size of the patterned structures.The UV-irradiated material has the same optical index as the cyanine dye film, but the Lorentzian peak related to the sharp absorption at 2.1 eV is eliminated (see Supplementary Materials). Optical properties of the samples are studied using leakage radiation microscopy\cite{Drezet}through a Nikon oil immersion objective of 1.49 numerical aperture (a more detailed description of the experimental setup can be found in \cite{Aberra}). Reflectometry images in the Fourier plane, which represent the dispersion relation (energy as a function of wavevector), are collected by imaging the microscope objective Fourier plane by a Nikon CCD camera or an Andor spectrometer, coupled to a CCD camera. Luminescence is triggered with a 532 nm laser diode beam focused onto the upper face (dye layer) of the sample. The emitted light is collected in direct space or Fourier space by a CCD camera or a spectrometer. Direct space images, with selected region of Fourier space can be performed by mounting an iris (selecting the central part of the Fourier plane) or a circular beam blocker in an intermediate Fourier plane.

Before addressing the patterned surfaces, the properties of uniform layers of UV-irradiated and of active dye on silver were measured. Figure 1(a) shows the dispersion relation of irradiated dye on silver. The line corresponding to the surface plasmon is uniformly broadened showing that the dye absorption resonance has been completely removed by the irradiation. On the other hand, the dispersion relation of the active dye film on silver (Fig. 1(b)) shows a typical polariton anticrossing associated with the exciton/plasmon hybridization in a strong coupling regime. The Rabi splitting at resonance is 155 meV with a bare dye broadening of 30 meV and a plasmon linewidth of 40 meV, fulfilling the strong coupling conditions\cite{Torma}. The polariton coherence length deduced from its wavevector width at resonance is about 7 $\mu$m\cite{Chevrier}(see Supplementary Materials).

\begin{figure} \includegraphics[width=8.5cm]{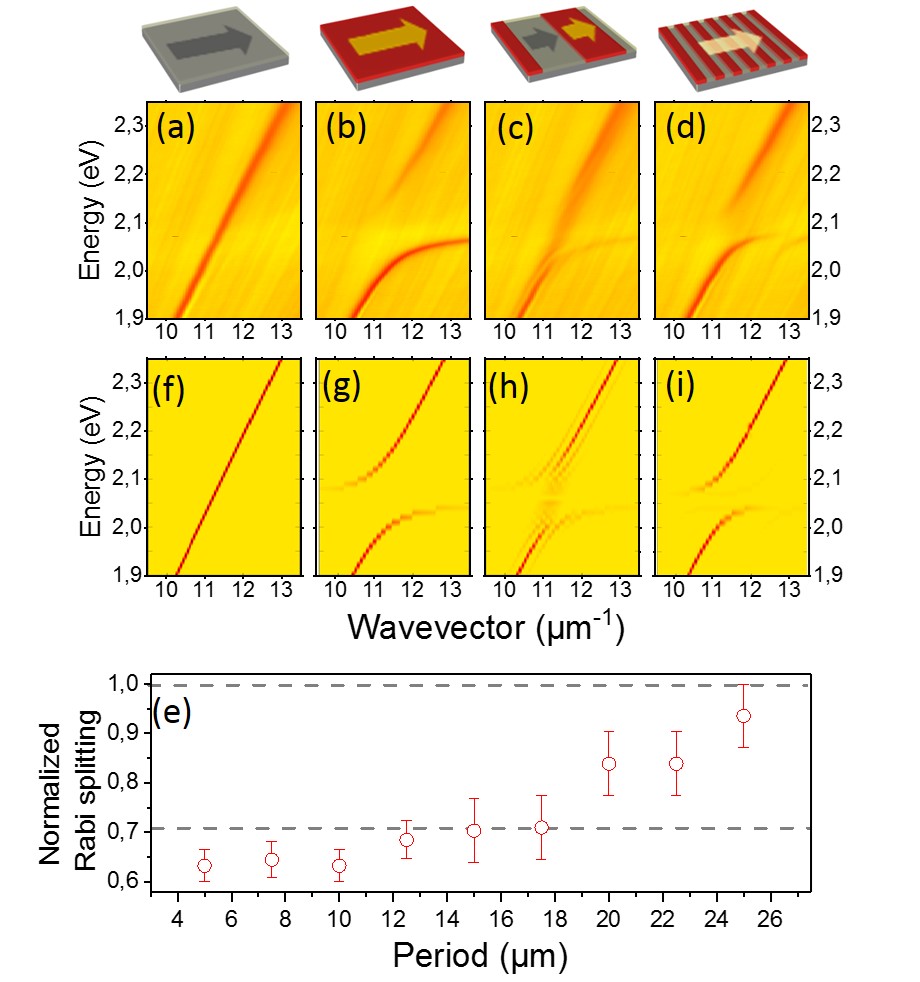}
\caption{(a-d) Experimental reflectometry dispersion images of a 50 nm silver film covered by a 17 nm thick dye layer, measured along the propagation direction perpendicular to the stripes: (a) homogeneously UV-irradiated film; (b) homogeneous active dye film; (c) patterned layer with a large period (25 $\mu$m); (d) patterned layer with small period (5 $\mu$m). (e) Rabi splitting recorded as a function of the array period and normalized to the Rabi splitting of a homogeneous active dye layer. The dashed lines correspond to a Rabi splitting decreased by a factor of $ \sqrt{2} $ compared to a full active dye film. (f-i) Computed dispersion curves for continuous and patterned films: (f) bare plasmon dispersion; (g) plasmon coupled to a continuous active dye film; (h) patterns of  25 $\mu$m period; (i) patterns of 5 $\mu$m period. All the patterned samples have a filling factor of 1/2.}
\end{figure}

To demonstrate averaging of optical properties within the coherence length, characteristic of metasurfaces, we compared patterned films with two periods: one larger (25 $\mu$m) and the other smaller (5 $\mu$m) than the coherence length. In both cases the filling factor was 1/2. The dispersion relations recorded for propagation direction (wavevector) perpendicular to the stripes are shown in Figs. 1(c) and 1(d). For the larger period (Fig. 1(c)), the dispersion shows the coexistence of a continuous line and an anticrossing between two branches. The Rabi splitting associated with the anticrossing is 145 meV, which is close to the one obtained with the homogeneous dye film (Fig. 1(b)), while the continuous line has essentially the same position as the bare plasmon line (Fig. 1(a)). This behavior can be viewed as the response of two distinct regions with strong and weak plasmon-molecule coupling, and negligible interaction between them. For the smaller period (Fig. 1(d)), the measured dispersion relation is drastically different: only an anticrossing is visible, showing no regular plasmon propagation anymore. Furthermore, the Rabi splitting at resonance ($ \hbar\Omega $= 98 meV) is reduced compared to the continuous dye film, confirming the effective behavior of the metal/dielectric metasurface, in that it differs from the basis constituents. It has to be noticed that another anticrossing, translated by  $k=$1.31$\mu$m$^{-1}$, can be seen in Fig.1(d). This additional line corresponds to the first diffraction order of the periodic structure (for the 5$\mu$m period, 2$\pi$/period=1.26$\mu$m$^{-1}$) and is also present for larger periods.

The first qualitative interpretation for these observations is that, for small periods, the system averages properties of the materials with and without the dye resonance absorption. Since the average number of emitters interacting with the plasmon mode is now reduced by a factor of 2 compared to the continuous dye layer, the effective interaction energy should be reduced by a factor $ \sqrt{2} $\cite{Bonnand}. Patterns of different periods were studied and the extracted Rabi splittings at resonance are summarized in Fig. 1(e). The splittings have been normalized with respect to the Rabi splitting of the continuous dye film. For small periods up to 15 $\mu$m, a quasi-plateau can be measured, while for large periods the splitting reaches the continuous film value. The striking point lies in the large value of the period at which the transition between the mean-field (effective) and multicomponent (polariton and plasmon) behavior occurs, more than one order of magnitude greater than the wavelength.\\

The properties of the system experimentally studied above can be theoretically rationalized by invoking a quantum mechanical framework for strong plasmon-exciton coupling. The resulting Hamiltonian (see Supplementary Material for details) is given by $ H=\sum_{K}H_{K} (\hbar=1) $, where
\[H_{K}=\sum_{\kappa}^{}\omega_{K+\kappa}a^{\dagger}_{K+\kappa}a_{K+\kappa}+\omega_{e}\sum_{\kappa'}^{\prime}\sigma^{\dagger}_{K+\kappa^{'}}\sigma_{K+\kappa^{'}}\]
\vspace{-0.7cm}
\[+\sum_{\kappa}^{}\sum_{\kappa'}^{\prime}\left[J_{K+\kappa}K_{\kappa-\kappa'}\sigma^{\dagger}_{K+\kappa'}a_{K+\kappa}+h.c.\right] ...  (1)\]

Here, the two-dimensional wavevector K belongs to the Brillouin zone $ \left[0,\frac{2\pi}{L}\right) \times\left[\dfrac{-\pi}{a},\dfrac{\pi}{a}\right) $ defined by the unit cell of the simulation, taken to be $L\times a$, where $a\ll L$ is the smallest spatial resolution of the simulation and \textit{L} is the patterning period. $a_{k}^{\dagger}\left(a_{k}\right)$ creates (annihilates) a plasmon with energy $\omega_{k}$, at wavevector $k=$.  $\sigma_{k}^{\dagger}\left(\sigma_{k}\right)$ is the active dye Frenkel exciton operator in Fourier space, which creates (annihilates) a delocalized collective exciton with energy $\omega_{e}$ at wavevector $k=$ (see supplementary material). $ J_{k} $ is a collective plasmon-exciton coupling for the continuous dye film (with no patterning). Notice that the dye patterning enables mixing between plasmons and excitons (created by the operator $\sigma_{k}^{\dagger}$) as long as their wavevectors differ by a Bragg vector $ \kappa=\kappa_{x}\widehat{x} $, where $\kappa_{x}=\dfrac{2\pi m}{L}  $(with $ m $ integer). The efficiency with which this mixing occurs is dictated by the Fourier transform $ F_{\Delta\kappa}=\dfrac{1}{\zeta}\sum_{\nu=0}^{N_{uc}-1}p_{\nu}\exp(i\Delta\kappa\upsilon a) $  of the patterning function $ p_{\nu} $, where $ \nu a=0,a,\cdots,(N_{uc}-1)a $ are the $x-$positions of the localized molecular excitations (the stripes are along the \textit{y} axis, the patterning along \textit{x}) and $N_{uc}a=L $. Given a filling factor $f$, $ p_{\nu}=1 $ for $ \nu=0,\cdots,fN_{uc}-1 $ and $ p_{\nu}=0 $ otherwise, while the normalization is given by $ \zeta=\sqrt{f}N_{uc} $. Finally, $ \sum_{\kappa'}^{'} $ indicates a constrained summation over Bragg vectors associated with an integer number of wavelengths per active stripe; for half-filling factor, this subset is given by $ \kappa'=\left( \kappa'_{x},0\right)  $ with $ \kappa'_{x}=\dfrac{2\pi m'}{L} $ for $ m' $ even. It is clear that in the absence of patterning, $ F_{\kappa-\kappa'}=\delta_{\kappa,\kappa'} $ and Eq. (1) reduces to the standard polaritonic dispersion Hamiltonian.

Plasmon-weighed dispersion plots are obtained by numerical diagonalization of $ H_{K} $(see Supplementary Materials). Results for the continuous inactive and active dye films are plotted in Figs. 1(f) and 1(g), respectively. For a patterned film of small period (5 $\mu$m, Fig. 1(i)), a very clear anticrossing with $ 1/\sqrt{2} $ Rabi splitting of the continuous film can be observed, verifying our effective medium picture. For a large period of 25 $\mu$m (Fig. 1(h)), the stripes support a large number of modes, upon which the calculation shows a clear partitioning between those which are confined in the active and inactive dye stripes, respectively. The former feature strong coupling with the dyes, while the latter remains as plasmons localized in the inactive dye stripes. As no losses are included in the calculation, the calculated lines have no width. The losses, which limit the coherence length, broaden the experimental lines and explain the merging of the lines which is observed in the experiments: the  plasmonic discrete modes appearing in the anticrossing gap form the propagating plasmon band, and the different diffracted polaritonic lines tend to merge to form a new anticrossing.

Another feature of the metasurface geometry studied here is its anisotropy, which has been investigated for a small pattern period (5 $\mu$m) by recording the dispersion relations for different propagation directions. For propagation along the stripes, the dispersion presented in Fig. 2(a) can be understood as the coexistence of bare plasmon and polariton lines, in contrast to that recorded perpendicular to the stripes (Fig. 2(b)) which only shows polariton lines. The value of the Rabi splitting parallel to the stripes is 160 meV, almost equal to that of the continuous dye film. A first interpretation based on this energy is that the effective medium effect is not present for propagation parallel to the stripe: polaritons confined along the active dye stripes coexist with propagating plasmons confined along the bare metal regions. Our numerical calculations unambiguously reproduce the anisotropic dispersion of the system as shown in .Figs. 2(c) and 2(d). Figure 2(e) shows the full experimental two-dimensional dispersion for a 5 $\mu$m period. This surface is deduced by tracking reflectivity minima as a function of azimuthal angle. A propagating plasmon is present only for angles between $ \pm20° $ about the direction of the stripes. For larger angles, the bare plasmon line is not detected anymore, and only the polariton lines remain.

\begin{figure}
\includegraphics[width=9.0cm]{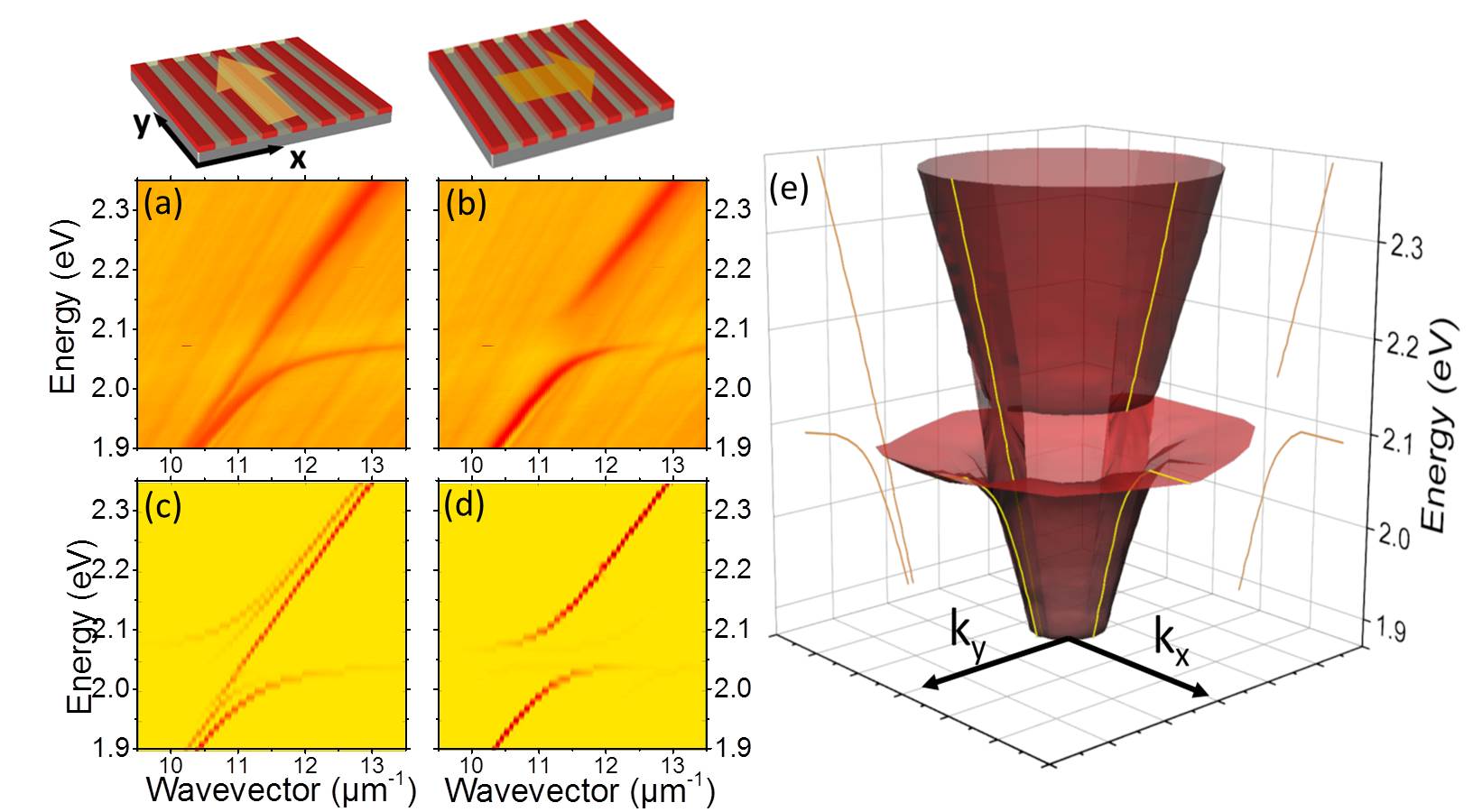}
\caption{Dispersion relations measured for a 5 $\mu$m period pattern with propagation direction (a) parallel to the stripes and (b) perpendicular to the stripes, as well as corresponding calculated dispersion relations (c),(d). (e) Two-dimensional experimental dispersion surface reconstructed across all propagation directions (from directions orthonormal to the stripes $k_{x} $ to directions parallel to the stripes $k_{y} $). For clarity, the region of wavevectors with magnitudes lower than 10 $\mu m^{-1}$ is removed. }
\end{figure}

The described metasurfaces are also accompanied by substantial modifications in the emission properties of the constitutive dye. The hybridization of a large number of molecular excited states with an extended plasmonic mode results in two types of states: extended-polaritonic states (with a well-defined wavevector) and localized incoherent-dispersionless purely molecular states also known as dark or reservoir states\cite{Agranovich,Ribeiro}. The polaritonic states and the plasmon modes can be respectively selected by choosing wavevector ranges under the light cone (non-radiative modes). Figure 3(a) presents a Fourier image of the emission of a polaritonic metasurface, with the horizontal axis $x$ perpendicular to the stripes, non-resonantly excited by a continuous 532 nm laser. Figure 3(b) shows the direct luminescence imaging of the incoherent states, where the emission coupled to the plasmon is removed by selecting the central part of the Fourier plane, i.e. the small wavevectors. In this case, the dye stripes appear clearly, and no emission is observed between them in the inactive dye areas.

\begin{figure}
 \includegraphics[width=8.5cm]{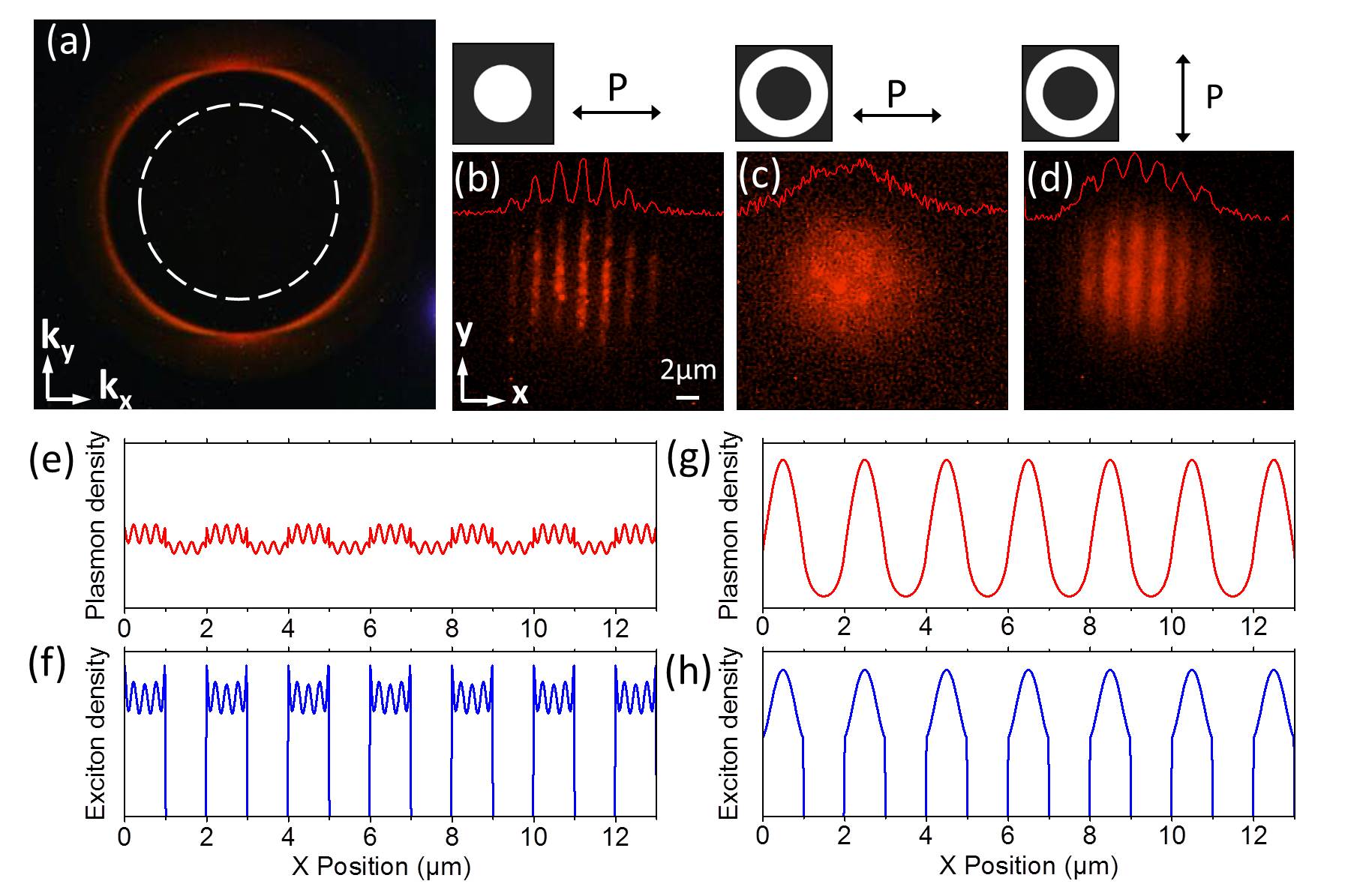}
\caption{Emission properties of polaritonic metasurface with a 2 $\mu$m period, with stripes oriented along the y direction. (a) Fourier space image of the luminescence. The white dashed circle limits the Fourier space filtering areas used in panels (b-d) for direct space imaging. (b) Direct emission image of the incoherent states. (c-d) Direct emission image associated with large wavevectors with a polarization direction perpendicular (c) and parallel (d) to the stripes. The intensity profile at the center of the image along the $x$ axis is plotted as a red line on top of each image. (e-h) Plasmon and exciton density of the calculated low energy polariton wavefunction at resonance for propagation perpendicular (e),(f) and parallel (g),(h) to the stripes.}
\end{figure}

When the plasmon/polariton emission is selected (large wavevectors), a strong polarization dependence can be observed. Selecting the polarization in real space is equivalent to choosing the propagation direction in reciprocal (Fourier) space, as the plasmon and thus the polaritons are TM polarized. Strikingly, for a horizontal polarization (i.e. propagation direction $ k_{x} $, Fig. 3(c)) the stripes are hardly visible. This uniform luminescence confirms the interpretation of an effective medium behavior for periods smaller than the coherence length. For a vertical polarization (i.e. propagation direction  $ k_{y} $, Fig. 3(d)) the stripes appear clearly although with a lower contrast than the incoherent luminescence. To analyze the luminescence images, the low energy polariton wavefunction at the resonance has been calculated. Figures 3(e) and 3(f) shows the spatial partition of the plasmonic and excitonic densities associated with this wavefunction. An essentially uniform plasmonic density is obtained for a propagation perpendicular to the stripes whereas large maxima within the stripes can be seen for a propagation parallel to the stripes. As the intensity collected in the experiments is associated with the plasmonic part\cite{Norris} of the wavefunction, the calculated densities and measured luminescence are fully consistent.

\begin{figure}
 \includegraphics[width=8 cm]{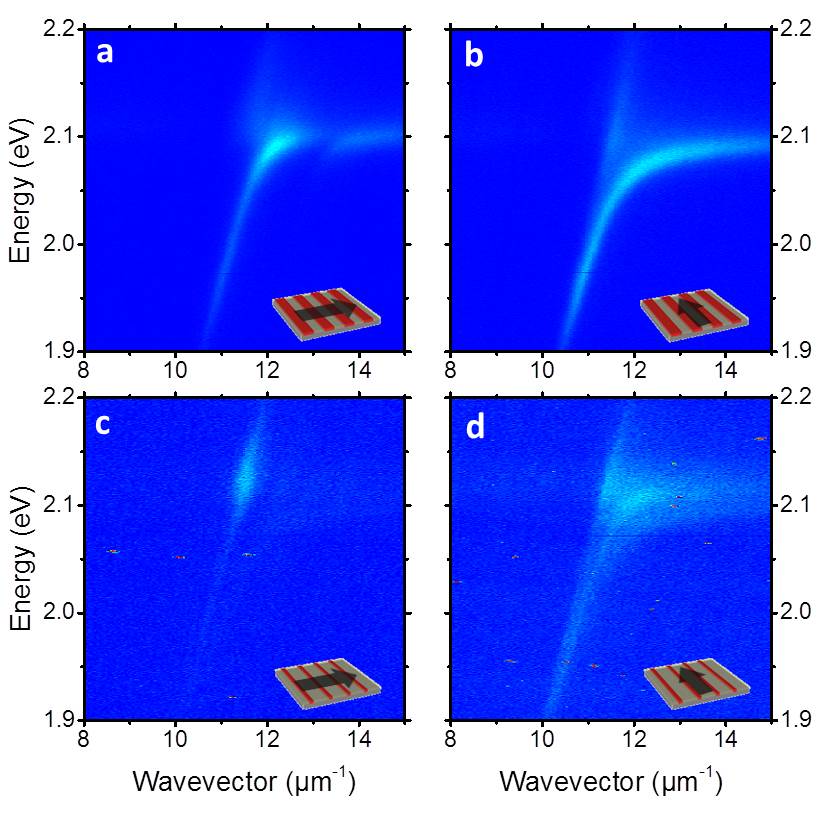}
\caption{Anisotropic emission dispersion properties of polaritonic metasurfaces with two different filling factors. (a-b) Filling factor 1/2 (stripe width is 2.5 $\mu$m, period is 5 $\mu$m). (c-d) Filling factor 1/10 (stripe width 1 $\mu$m).}
\end{figure}

The hybridization with the plasmon renders luminescence dispersive and modifies emission energies. The luminescence as a function of wavevector for the 5 $\mu$m period array (dispersion of Fig. 2) is shown in Figs. 4(a) and 4(b). The polariton appears for both propagation directions, while an emission through the plasmon can be detected only along $ k_{y} $. By changing the filling factor, the effective concentration of active dye molecules within the plasmon coherence length can be modified, allowing a transition from strong to weak coupling. With a filling factor of 1/10, an emission without any polaritonic contribution can be observed on Fig. 4(c) for a detection perpendicular to the stripes. This emission corresponds to the typical weak coupling regime for molecules close to a metal surface: an emission at the dye transition energy into the plasmon. For propagation along $ k_{y} $ (Fig. 4(d)), both emission into plasmon and a polariton can be detected. These observations agree with the previous interpretation of two coexisting modes, spatially localized in the regions of the active (polariton) and inactive (plasmon) dyes. We have thus demonstrated unidirectional polaritons and anisotropic luminescence starting from an isotropic emitting material.

In conclusion, we have shown that the properties of molecules strongly coupled to surface plasmons can be drastically modified by structuring the sample on distances which are large compared to wavelengths. The modified material band structure can find applications in polarization controlled sources, as well as in topological photonics based on anisotropic dispersions and Dirac cones\cite{Zhou} or exceptional points\cite{Gao}. This control is reinforced by the large variety of 2D lattices that can be patterned in the active material, associated with the simplified fabrication of the supra-wavelength features required for polariton metasurfaces. The anisotropic behavior demonstrated here could also open the way to anisotropic conductivity based on strong coupling\cite{Orgiu}. The active polaritonic metasurfaces we propose are not restricted to plasmonic modes nor organic dyes alone, but can be extended to a large variety of structures in strong coupling such as cavity polaritons which have similar extended coherence lengths\cite{Shalabney}. Finally, building on the excitonic part of the hybrid states, these ideas can also be harnessed to design novel ultrafast reconfigurable metasurfaces which can be optically\cite{Gunter,Ditinger,Pacifici}  or electrically\cite{Anappara} switched upon pumping.

S.K.S. thanks the Center for Excitonics, an Energy Frontier Research Center funded by the U.S. Department of Energy, under Award No. DE-SC0001088 and also the Ministry of Education and Science of the Russian Federation for supporting the research in the framework of the state assignment, Award No. 3.2166.2017/4.6. J.Y.Z. acknowledges support from NSF CAREER Award CHE-164732.

\end{document}